\documentclass[aps,prl,twocolumn,showpacs,eqsecnum]{revtex4}
\usepackage{graphicx}
\usepackage{epstopdf}

\def\lesssim{\mathrel{\rlap{\lower4pt\hbox{\hskip1pt$\sim$}}}<}
\def\gtrsim{\mathrel{\rlap{\lower4pt\hbox{\hskip1pt$\sim$}}}>}

\begin{document}

\title{Dark matter and the first stars: a new phase of stellar evolution}

\author{
\mbox{Douglas Spolyar{$^1$},}
\mbox{Katherine Freese{$^{2,3}$},}
and
\mbox{Paolo Gondolo{$^{4}$}}
}

\affiliation{
\mbox{$^1$ Physics Dept., University of California, Santa Cruz, CA 95064} 
\mbox{$^2$ Michigan Center for Theoretical Physics, 
Dept. of Physics, University of Michigan, Ann Arbor, MI 48109}
\mbox{$^3$ Visiting Miller Professor, Miller Institute, 
University of California, Berkeley, CA 94720}
\mbox{$^4$ Physics Dept., University of Utah, Salt Lake City, UT 84112}
\\
{\tt dspolyar@physics.ucsc.edu},
{\tt ktfreese@umich.edu},
{\tt paolo@physics.utah.edu}
}

\begin{abstract} \noindent
A mechanism is identified whereby dark matter (DM) in protostellar
halos dramatically alters the current theoretical framework for the
formation of the first stars.  Heat from neutralino DM annihilation is
shown to overwhelm any cooling mechanism, consequently impeding the
star formation process and possibly leading to a new stellar phase.  A
``dark star'' may result: a giant ($\gtrsim 1$ AU) hydrogen-helium
star powered by DM annihilation instead of nuclear fusion.
Observational consequences are discussed.
\end{abstract}

\pacs{97.10.Bt,95.35.+d,98.80.Cq}

\maketitle

The first stars in the universe mark the end of the cosmic dark ages,
reionize the universe, and provide the enriched gas required for later
stellar generations.  They may also be important as precursors to
black holes that coalesce and power bright early quasars.  The first
stars are thought to form inside haloes of dark matter of mass $ 10^5
- 10^6 M_\odot$ at redshifts $z=10-50$.  These haloes arose from the
merging of smaller structures, as overdense regions in the universe
assemble heirarchically into ever larger haloes. The haloes consist of
85\% dark matter and 15\% baryons in the form of pristine hydrogen and
helium (from Big Bang nucleosynthesis).  The baryonic matter cools and
collapses via molecular hydrogen cooling \cite{Hollenbach} into a
single small protostar \cite{omukai} at the center of the halo (for
reviews see e.g. \cite{Ripamonti:2005ri,Barkana:2000fd,Bromm:2003vv}).

In this paper we consider the effect of the dark matter particles on the
formation process of these first stars.  We focus on the most
compelling dark matter candidate, the lightest supersymmetric
particle. Supersymmetry (SUSY), at this point a beautiful theoretical
construct, has the capability of addressing many unanswered questions
in particle theory as well as providing the underpinnings of a more
fundamental theory such as string theory.  If SUSY is right, then for
every known particle in the universe, there is an as yet undiscovered
partner. The lightest of these, the Lightest Supersymmetric Particle
or LSP, would provide the dark matter in the universe.  The search for
SUSY is one of the motivations for building the Large Hadron Collider
at CERN, and one may hope that it will be discovered as early as 2008.

The LSP is the favorite dark matter candidate of many physicists.
This is true not only because of the beautiful properties of SUSY, but
also because the LSP automatically has the right properties to provide
24\% of the energy density of the universe.  In particular, the
neutralino, the SUSY partner of the W, Z, and Higgs bosons, has the
required weak interaction cross section and $\sim$ GeV - TeV mass to
give the correct amount of dark matter in the universe today.  The
SUSY particles are in thermal equilibrium in the early universe, and
annihilate among themselves to produce the relic density today.  It is
this same annihilation process that is the basis of the work we
consider here.  The SUSY particles, also known as WIMPs (Weakly
Interacting Massive Particles), annihilate with one another wherever
their density is high enough.  Such high densities are achieved in the
early universe, in galactic haloes today \cite{ellis,gs}, in the Sun
\cite{sos} and Earth \cite{freese,ksw}, and, as we will show, also in
the first stars. As our canonical values, we will use the
standard value $\langle \sigma v \rangle = 3 \times 10^{-26} {\rm
cm^3/sec}$ for the annihilation cross section and $m_\chi = 100$ GeV
for the SUSY particle mass; but will also consider a broader range of
WIMP masses (1 GeV--10 TeV) and cross-sections \footnote{ The details
of the interactions and masses of the neutralinos depend on a large
number of model parameters. In the minimal supergravity model,
experimental and observational bounds restrict the neutralino mass
$m_\chi$ to 50 GeV--2 TeV, while the annihilation cross section
$\sigma v$ lies within an order of magnitude of $\langle \sigma v
\rangle = 3 \times 10^{-26} {\rm cm^3/sec}$ (except at the low end of
the mass range where it could be several orders of magnitude smaller).
Nonthermal particles can have annihilation cross-sections that are
many orders of magnitude larger (e.g.~\cite{moroi}) and would have
even more drastic effects.}.  The effects we find apply equally well
to other WIMP candidates, such as sterile neutrinos or Kaluza-Klein
particles.

\begin{figure}[t]
\centerline{\includegraphics[width=0.5\textwidth]{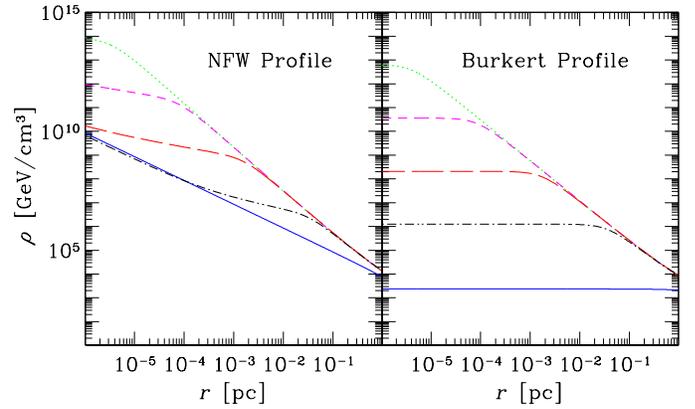}}
\caption{ Adiabatically contracted DM profiles for (a) an initial NFW
profile and (b) an initial Burkert profile, for $M_{\rm vir}=10^6
M_\odot$, $c_{\rm vir}=10$, and $z=19$.  The blue (solid) lines show
the initial profile.  Black (dot-dash) lines correspond to a baryonic
core density of $10^7{\rm cm}^{-3}$, red (long-dashed) lines to
$10^{10}{\rm cm}^{-3}$, magenta (dashed) lines to $10^{13}{\rm
cm}^{-3}$, and green (dotted) lines to $n\sim 10^{16}{\rm cm}^{-3}$.
\vspace{-\baselineskip}
}
\end{figure}

{\it DM density profile.\/} In order to study the effects of DM on the
star formation process, we need to know its density profile inside the
baryonic core.  While simulations have obtained remarkably good
density profiles for the collapsing protostellar gas, they have
unfortunately (as yet) been unable to do so for the DM.  Thus we use
adiabatic contraction \cite{Blumenthal:1985qy} to obtain estimates of
the DM profile.  As our initial conditions, we take an overdense
region of $10^5-10^6 M_\odot$ with an NFW profile \cite{NFW} for both
DM and gas, where the gas contribution is 15\% of that of the
DM. (For comparison, we also use a Burkert profile \cite{burkert}, which has a DM
core before contraction.) Then, as the gas collapses, we allow the DM to respond to the
changing baryonic gravitational potential (gas density profiles taken
from simulations of \cite{ABN,Gao06}). The final DM density profiles
are computed with adiabatic contraction ($M(r)r$ = constant) and are
shown in Fig.~1 for concentration parameter $c=10$ at a redshift
$z=19$ and halo mass $M=10^6 M_\odot$.  It is important to point out
that our results do not change much for other choices of these
parameters; e.g. even for $c=1$, the dark matter density only changes
by a factor of 4.  After contraction, the DM density at the outer edge
of the baryonic core is roughly $ \rho_\chi \simeq 5 {\rm GeV/cm}^{-3}
(n/{\rm cm}^{3})^{0.81} $ and scales as $\rho_\chi \propto r^{-1.9}$
outside the core.

Our adiabatically contracted NFW profiles match the DM profile
obtained numerically in~\cite{ABN}
(see their Fig.2).
They present two DM profiles (their earliest and latest profiles), for
$n \sim 10^3 {\rm cm}^{-3}$ and $n \sim 10^{13} {\rm cm}^{-3}$ as far
inward as $5 \times 10^{-3}$pc and $0.1$pc.  The slope of these two
curves is the same as ours.
$\rho_\chi \propto r^{-1.9}$.  If one uses this slope and extrapolates
inward to smaller radii and to higher densities, then one obtains the
same DM densities as with our adiabatic contraction approach. We are
encouraged by this agreement.  It is interesting to note that adiabatically contracted
objects of any mass, even outside of the context of Pop III stars,
might lead to enhanced DM annihilation leading to photons observable
by GLAST or to a modified reionization history.

As a caveat, we note that the approach of adiabatic contraction must
be used with caution.  It formally requires the orbital particle time
to be short compared to the collapse time, though in practice the
method works well beyond this limit (see e.g.~\cite{Steigman}).  We
are also concerned about the requirement of spherical symmetry, when
in fact there are filaments, clumps, and mergers, so that dynamical
friction or violent relaxation may take place.  The gas may form bars
that could sweep out DM, though the rotation timescale for bars may be
too long for them to be important.  We encourage simulators to improve
the DM resolution in first stars to confirm our results.  The closest
previous work is that of Merritt \cite{merritt}, who used initial
profiles $\rho \propto r^{-\gamma}$ with $\gamma = [0,2]$ around a
central black hole and found final profiles $\rho \propto
r^{-\gamma'}$ with $\gamma'=[2.25,2.5]$ (i.e. even steeper than we
what we found).  We caution the reader that our heating estimates
below rely upon DM densities obtained with these assumptions.

{\it DM Heating.\/} 
WIMP annihilation produces energy at a rate per unit volume $Q_{\rm
ann} = \langle \sigma v \rangle \rho_\chi^2/m_\chi \linebreak \simeq  1.2
\times  10^{-29} {\rm erg/cm^3/s} \,\,\, (\langle \sigma v \rangle / (3
\times 10^{-26} {\rm cm^3/s}))  \linebreak(n/{\rm cm^{-3}})^{1.62} (m_\chi/(100
{\rm GeV}))^{-1}$. In the early stages of Pop III star formation, when
the gas density is low, most of this energy is radiated away. However,
as the gas collapses and its density increases, a substantial fraction
$f_Q$ of the annihilation energy is deposited into the gas, heating it
up at a rate $f_Q Q_{\rm ann}$ per unit volume.

Previous work \cite{Ripamonti:2006gr,Chen:2003gz} considered the
effects of DM annihilation on earlier low density phases of Pop III
star formation ($n \lesssim 10^4 {\rm cm}^{-3}$).  They rightly
concluded that 100 GeV neutralinos cannot heat the protostar at these
low densities because their annihilation products simply escape out of
the object without depositing much energy inside.  They consequently
focused on lighter particles, such as 1--10 keV sterile neutrinos and
1--100 MeV light dark matter.

However, we find that, for WIMP mass $m_\chi = 100$GeV (1 GeV), a
crucial transition takes place when the gas density reaches $n>
10^{13} {\rm cm}^{-3}$ ($n>10^9 {\rm cm}^{-3}$).  Above this density,
most of the annihilation energy remains inside the core and heats it
up to the point where further collapse of the core becomes difficult.
At this point the DM density is 2\% (10\%) of the gas density in the
core, the size of the core is $\sim 17$ A.U.  ($\sim 960$ A.U.), its
mass is $\sim 0.6 M_\odot$ ($\sim 11 M_\odot$), $f_Q \sim 2/3$, and
the energy produced by DM heating is $\sim 140 L_\odot (\sim 835
L_\odot)$.

Here we estimate the fraction $f_Q$ of DM annihilation energy that
remains inside the gas core (more detailed work is in progress).
$f_Q$ scales linearly with the gas density and depends on the relative
number of the various annihilation products and their energy
spectrum. The latter are heavily dependent on the WIMP model. From our
experience with neutralino DM, we assume the following typical values:
about 30\% of the energy goes into neutrinos, 30\% into photons, and
30\% into stable charged particles like electrons and
positrons. Unstable particles like neutral pions, charged pions, and
muons decay into photons, neutrinos, and electrons before exiting the
cloud. The energy spectrum of photons and electrons depends to some
extent on the exact annihilation channels.  For our purpose, we
consider typical spectra produced in Pythia simulations of 500 GeV
neutralino annihilation \cite{DarkSUSY,Fornengo}. Other spectral
shapes will change the precise values of our results but not the
overall effect.

Neutrinos escape from the cloud without depositing an appreciable
amount of energy. Electrons below a critical energy $E_c \approx 280$
MeV in hydrogen lose energy predominantly by ionization. Higher energy
electrons do so by emission of bremsstrahlung photons. As these
bremsstrahlung photons pass near gas nuclei, they create electrons and
positrons, which in turn may generate other bremsstrahlung
photons. Thus starts a sequence of photon, electron, and positron
conversions: an electromagnetic (EM) cascade. While photons above
$\approx 100$ MeV also initiate an EM cascade, lower
energy photons transfer most of their energy to electrons in the gas
(Compton scattering).

We approximate the energy loss of electrons via ionization with 
$4.41{\rm MeV}/E $ (g/cm$^2$)$^{-1}$. For EM cascades, we 
assume a gamma distribution for the mean longitudinal profile. 
Thus the fraction of energy lost in traversing a thickness $X$ of gas 
equals $\gamma(a,bX/X_0)/\Gamma(a)$, where $\gamma(x,y)$ is 
the incomplete gamma function, 
$X_0=63$ g/cm$^2$ is the radiation length in hydrogen,
$a=1+b[\ln(E/E_c)-0.5]$, and $b=0.5$ \cite{rpp}.
We estimate the core thickness as $X=1.2 m_p n r_0$. Here 
$m_p$ is the proton mass, $r_0$ is the core radius, and the factor of 1.2 is 
appropriate for a uniform sphere.
We model the fraction of energy loss of photons by
converting each photon to an electron of the same energy after one
photon attenuation length. The latter is computed from formulas 
in \cite{rossi}, interpolated to produce the hydrogen curve in 
\cite{rpp}, figure 27.16.

Annihilation of DM outside the core can also 
contribute to heating inside the core. The
annihilation products are not stopped until they reach the core
because the baryon density outside falls as $r^{-2.3}$. 
We integrate the effects of 
DM particles annihilating in the external region
at $r>r_0$, whose annihilation products have a
probability of hitting the core given by 
$\pi r_0^2 / 4 \pi r^2$.  
Taking $\rho_\chi \sim 1/r^2$ outside the core, we find 
a 25\% enhancement in core heating due to the external 
region. Henceforth
our results are obtained without this enhancement factor.
 
{\it Results.\/} 
To compare with DM heating, we include all relevant cooling mechanisms.
The dominant mechanism is 
H$_2$ cooling (we use the rates in~\cite{Hollenbach}), 
though our code also includes other effects such as
H line cooling (important only at high
temperatures) and Compton cooling (important only if the gas becomes
very ionized) \cite{Tegmark:1996yt}. 
We use the opacities from \cite{Yoshida06}; e.g. at $n \sim 10^{13} {\rm
cm}^{-3}$, we take a $\sim 8\%$ cooling efficiency.
Setting the heating rate equal to the cooling rate gives the critical
temperature $T_c(n)$ at a given density $n$ below which heating dominates.  

In figure 2 we compare $T_c(n)$  to
typical evolution tracks 
in the temperature-density phase plane.  The blue (solid) and green 
(dotted) lines show
the temperature evolution of the protostellar gas in the
simulations (without DM) of
\cite{Yoshida06} and \cite{Gao06} respectively.
The red (dashed and dot-dashed) lines show the critical temperature: 
below these lines, DM heating dominates over all cooling mechanisms.  

\begin{figure}[t]
\centerline{\includegraphics[width=0.5\textwidth]{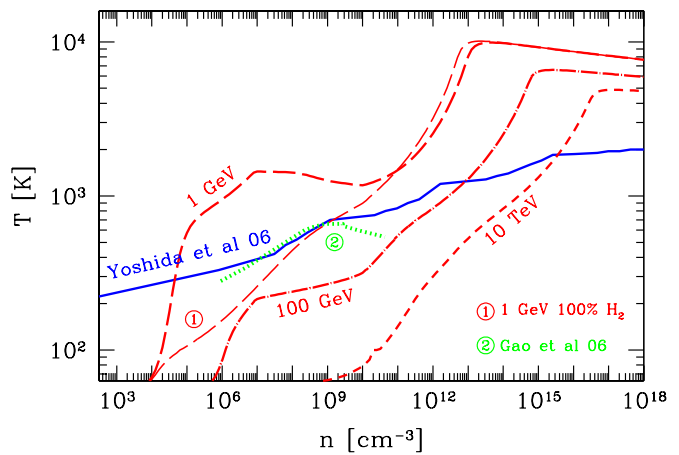}}
\caption{ Comparison of critical temperature  (red dashed lines) to
typical evolution tracks
in the temperature-density phase plane.  The blue (solid) and green 
(dotted) lines show the protostellar gas evolution from
simulations of
\cite{Yoshida06} and \cite{Gao06} respectively.
The red dashed lines mark $T_c(n)$ 
for: (i) $m_\chi = 1$ GeV with H$_2$ density from
simulations, (ii) $m_\chi = 1$ GeV assuming 100\% H$_2$, 
(iii) $m_\chi = 100$ GeV and (iv) $m_\chi  = 10$ TeV.
At the crossing point of the blue/green and red 
lines, DM heating dominates over cooling in the
core's evolution.
\vspace{-0.5\baselineskip}
}
\end{figure}

Figure 2 illustrates results for a range of WIMP masses from 
1 GeV--10 TeV for a canonical
$3 \times 10^{-26} {\rm cm^3}/{\rm sec}$ annihilation cross-section.
Since the heating rate scales as $\langle \sigma v \rangle / m_\chi$,
these same curves equivalently  
apply to a variety of cross-sections for a given WIMP mass.

The important result is that the blue/green (evolutionary) and red 
(critical temperature)
lines always cross, regardless of WIMP mass or H$_2$ fraction: this is a 
robust result.  As soon as the core density reaches this crossing point,
the DM heating dominates inside the core and changes its
evolution.  
Notice that at $m_\chi=$1 GeV, 
the crossing point for small ${H_2}$ fraction is at low densities, around 
$n \sim 10^5 {\rm cm}^{-3}$, in agreement with \cite{Ripamonti:2006gr}.
If the ${H_2}$ fraction is increased, cooling dominates 
for a longer time, as expected, but not forever.

Our results were obtained for two possible values for the H$_2$ fraction: 
the value given by the simulations without DM, and 
the case of 100\% molecular hydrogen. The fully molecular case
produces the most effective cooling of the core and 
hence may underestimate the importance of DM heating.
DM annihilation gives rise to additional electrons
which could in principle produce more H$_2$ and go in the direction of
cooling the star; however, for 100 GeV mass WIMPs, the DM only becomes
important at such high gas densities that the core has already reached
a 100\% H$_2$ fraction and this effect is irrelevant. For lighter WIMPs,
the transition to DM heating may occur at a lower H$_2$ fraction and a different baryon density.

The crossing point is stable against temperature changes
at constant density and composition: if the temperature increases away
from the critical value, cooling
begins to dominate and drives the temperature back down to the critical value;
if the temperature decreases, heating dominates and lowers the temperature.  
The reason for this behavior is that the H$_2$ cooling rate increases
with temperature while the DM heating is independent of temperature.

As soon as the DM annihilation products are contained inside the
protostellar core, the heating dominates over the cooling. Hence, for
100 GeV (1 GeV) neutralino DM, once the gas density reaches a critical
value of $\sim 10^{13} {\rm cm}^{-3}$ ($10^9 {\rm cm}^{-3}$), the
heating rate from DM annihilation exceeds the rate of hydrogen
cooling. The protostellar core is {\it prevented from cooling and
collapsing} further.  

{\it Discussion.\/}
Our main conclusion is that the standard picture of Pop III star
formation is drastically modified by neutralino dark
matter annihilation inside the protostellar object.
The DM annihilation provides a heat source that exceeds any
cooling mechanism and thereby hinders the further collapse of the
protostar.  

We propose that a new type of object is created, a ``dark star''
supported by DM annihilation rather than fusion.  One question is how
long such a phase of stellar evolution lasts.  If such an object were
stable for a long time period, it would even be possible for these
dark stars to still exist today.  Dark stars could last as long as the
DM annihilation timescale, $\tau_e = m_\chi/(\rho_\chi \langle \sigma
v \rangle) $ $\simeq$ $0.6~ {\rm Gyr}$ $(n/10^{13} {\rm
cm}^{-3})^{-0.8} $ $ (m_\chi/100 {\rm GeV}) $ $(\sigma v/3 \times
10^{-26}{\rm cm}^3 {\rm s}^{-1})^{-1}$. For our canonical case, we
find $ \tau_e \sim 600$ Myr ($\sim 15$ Myr) for $n = 10^{13} {\rm
cm}^{-3}$ ($n = 10^{15} {\rm cm}^{-3}$).  By comparison, the entire
timescale for collapse (without taking into account DM annihilation)
is $\sim 1$ Myr at $z=50$ or 100 Myr at $z=15$; even for the more
recent episodes, the dynamical time at the high densities considered
here is very short ($<10^3$ yr).  However, after this DM annihilates
away, it is possible that the DM hole in the small central core can
fill in again, depending on the DM orbits at this stage.  DM further
out can also continue to heat the core.  On the other hand, as baryons
continue to accrete onto the protostar, it is possible that the
annihilation shuts down sooner.  The lifetime of the dark star phase
is crucial to addressing the question of the effects it has on the
universe.

Much further work is required to address the evolution of these
objects; we here briefly speculate as to possible outcomes.  One
possibility is that the dark star phase still persists today, so that
these objects never reach the main sequence.  A second possibility
would be a shorter dark star phase, during which the gas core is in a
state of quasi-hydrostatic equilibrium.  For our 100 GeV case, both
the surface gravity and the hydrostatic force per unit mass from the
temperature and pressure gradient are of the same order of magnitude,
$\sim 10^{-3} $ cm/s$^2$. Outer material would continue to accrete
onto the quasi-hydrostatic core \cite{stahler}, probably accompanied
by the formation of an accretion shock.  Alternatively, once the
core's central temperature reaches $\sim 10^6$ K, deuterium burning
and a pp chain may start. The star would then emit radiation and
contract in a Kelvin-Helmholtz timescale, until the central density
and pressure is high enough to start the CNO cycle. The star would
finally reach the zero-age main sequence. In this scenario, Pop III
star formation would be delayed rather than blocked.  A third
possibility would be that the core's contraction slows down as a
consequence of DM heating, and yet the core continues to contract
further.  Then DM heating would continue to dominate over available
cooling mechanisms at ever higher baryon densities.

The effects of a dark star phase of stellar evolution could be very
interesting.  The reionization of the IGM could be quite different, as
would be the production of the heavy elements required to form all
future generations of stars. We clearly need to recompute stellar
structure with this new heat source, to see how different dark stars
would look relative to the ordinary fusion-driven stars.  It is
possible, e.g., that dark stars are luminous but at lower
temperatures, so that UV radiation does reionize the IGM yet the
scenario is quite different. It is possible, e.g., that reionization
is delayed.  Perhaps the discrepancy between measurements of
$\sigma_8$ by WMAP3 and in Lyman-$\alpha$ could be resolved in this
way.

DM heating may also alter the mass of Pop III stars.  Even without DM
heating, the mass is still uncertain.  Refs.~\cite{mckee,Gao06} have
explored different possibilities for the accretion process (disk vs.\
spherical) of baryonic matter onto the central protostar as well as
sensitivity to cosmology and found great uncertainty as to the final
stellar mass.  DM heating could also affect the result.  For example,
in the case of spherical accretion, the heating might produce
radiation at the Eddington luminosity whose pressure prevents further
accretion.  Thus, the DM heating might lead to lower mass stars than
in the standard picture.  Alternatively the initial protostellar
object may be larger, and dark stars might accrete enough material
\cite{work} to form large black holes \cite{li,Pelupessy} en route to
building the (as yet unexplained) $10^9 M_\odot$ black holes observed
at $z \sim 6$.

What are other observational consequences of a ``dark star''?  If
these objects are luminous but differ from ordinary stars (e.g. shine
at lower temperatures), then James Webb Space Telescope could in
principle find them (at z$\sim$10) and differentiate their spectra
from those expected in the standard first star formation scenarios.
In addition, one might hope to detect the DM annihilation products
such as neutrinos and $\gamma$-rays.  However, the angular resolution
of current and planned detectors (AMANDA, ICECUBE, GLAST, HESS,
VERITAS, MAGIC) is not good enough to identify an individual dark star
source at $z> 10$, so that the $\nu$ and $\gamma$s would add to the
extragalactic backgrounds and could provide limits at best. Remnants
today of the dark star phase would be more testable, such as stellar
remnants still in enhanced DM distributions. Today's remnant $10^6$ DM
halos that were once the site of star formation may be modified due to
the adiabatic contraction that took place earlier, so that enhanced DM
annihilation might still occur today and these objects could be
identified in upcoming experiments as individual $\nu$ or $\gamma$
sources.

{\bf Acknowledgments.}  This paper would not have come into existence
without the help of Pierre Salati; we thank him for our conversations
at the inception of this project. We are also extremely grateful to
Chris McKee for his encouragement and for a series of discussions
during the progress of this work.  We thank A. Aguirre, P. Madau,
F. Palla, J. Primack, R. Schneider, S. Stahler, G. Starkman, and
N. Yoshida for discussions.  K.F. acknowledges support from the DOE
and MCTP via the Univ.\ of Michigan; from the Miller Inst.\ at UC
Berkeley; and thanks the Physics Dept.\ at UCSC.  D.S. and K.F. thank
the Galileo Galilei Inst. in Florence, Italy, for
support. P.G. acknowledges NSF grant PHY-0456825.  D.S. acknowledges
NSF grant AST-0507117 and GAANN.


\begin{thebibliography}{99}

\bibitem{Ripamonti:2005ri}
  E.~Ripamonti and T.~Abel,
  astro-ph/0507130.

\bibitem{Barkana:2000fd}
  R.~Barkana and A.~Loeb,
  Phys.\ Rept.\  {\bf 349}, 125 (2001).

\bibitem{Bromm:2003vv}
  V.~Bromm and R.~B.~Larson,
  Ann.\ Rev.\ Astron.\ Astrophys.\  {\bf 42}, 79 (2004).

\bibitem{ellis}
J.R. Eliis, R.A. Flores, K. Freese, S. Ritz, D. Seckel, and J Silk,
Phys.\ Lett.\ B {\bf 214}, 403 (1988).

\bibitem{gs}
  P. Gondolo and J. Silk, Phys.\ Rev.\ Lett.\ {\bf 83}, 1719 (1999).

\bibitem{sos}
  M. Srednicki, K.A. Olive, and J. Silk, Nucl.\ Phys.\ B {\bf 279},
  804 (1987).

\bibitem{freese}
  K.~Freese,
  Phys.\ Lett.\  B {\bf 167}, 295 (1986).

\bibitem{ksw} 
  L.M. Krauss, M. Srednicki, and F. Wilczek, Phys.\ Rev.\ D {\bf 33},
  2079 (1986).


\bibitem{Peebles:1968nf}
  P.~J.~E.~Peebles and R.~H.~Dicke,
  Astrophys.\ J.\  {\bf 154}, 891 (1968).

\bibitem{matsuda}
  T.~Matsuda, H.~Sato and H.~Takeda,
  Prog.\ Theor.\ Phys.\  {\bf 46}, 416 (1971).

\bibitem{Hollenbach}
  D.~Hollenbach and C.~F.~McKee,
   Astrophys.\ J.\  Suppl.\ {\bf 41}, 555 (1979).

\bibitem{moroi}
  T.~Moroi and L.~Randall,
  Nucl.\ Phys.\  B {\bf 570}, 455 (2000).
  
\bibitem{omukai}
  K.~Omukai and R. Nishi,
  Astrophys.J. {\bf 508}, 141 (1998).

\bibitem{NFW}
  J.~F.~Navarro, C.~S.~Frenk and S.~D.~M.~White,
  Astrophys.\ J.\  {\bf 462}, 563 (1996).

\bibitem{Blumenthal:1985qy}
  G.~R.~Blumenthal et al., 
  Astrophys.\ J.\  {\bf 301}, 27 (1986).

\bibitem{ABN}
  T.~Abel, G.~L.~Bryan and M.~L.~Norman,
  Science {\bf 295}, 93 (2002).

\bibitem{Gao06}
 L.~Gao et al., 
  astro-ph/0610174.
  
\bibitem{burkert}
  A.~Burkert,
  IAU Symp.\  {\bf 171}, 175 (1996)
  [Astrophys.\ J.\  {\bf 447}, L25 (1995)].

\bibitem{Steigman}
  G.~Steigman et al., 
  Astron. J. {\bf 83}, 1050 (1978).

\bibitem{merritt}
  D.~Merritt,
  arXiv:astro-ph/0301257.

\bibitem{Ripamonti:2006gr} 
  E.~Ripamonti, M.~Mapelli and A.~Ferrara,
  Mon.\ Not.\ Roy.\ Astron.\ Soc.\ {\bf 375}, 1399 (2007).

\bibitem{Chen:2003gz}
  X.~L.~Chen and M.~Kamionkowski,
  Phys.\ Rev.\  D {\bf 70}, 043502 (2004).

\bibitem{DarkSUSY} P.~Gondolo et al., J.\ Cosmol.\ Astropart.\ 
Phys.\ {\bf 7}, 8 (2004).

\bibitem{Fornengo} N.~Fornengo, L.~Pieri, and S.~Scopel, 
Phys.\ Rev.\ D {\bf 70}, 103529 (2004).
 
\bibitem{Kuhlen:2005es}
  M.~Kuhlen and P.~Madau,
  Mon.\ Not.\ Roy.\ Astron.\ Soc.\  {\bf 363}, 1069 (2005).

\bibitem{rpp} Particle Data Group: W.~M.~Yao et al.,
J.\ Phys.\ G {\bf 33}, 1 (2006).

\bibitem{rossi} B. Rossi, {\it High-Energy Particles} (Prentice-Hall, Inc., 
Englewood Cliffs, NJ, 1952).

\bibitem{Tegmark:1996yt}
  M.~Tegmark et al., 
F.~Palla,
  Astrophys.\ J.\  {\bf 474}, 1 (1997).

\bibitem{Yoshida06}
  N.~Yoshida et al., 
  Astrophys.\ J.\  {\bf 652}, 6 (2006).

\bibitem{work}
Work in progress.

\bibitem{stahler} S.~W.~Stahler, F.~Palla, and E.~E.~Salpeter, 
Astrophys.\ J.\ {\bf 308}, 697 (1986).

\bibitem{mckee} J.~C.~Tan and C.~F.~McKee,
  Astrophys.\ J.\  {\bf 603}, 383 (2004).
  
\bibitem{li}
  Y.~X.~Li {\it et al.},
  arXiv:astro-ph/0608190.

\bibitem{Pelupessy}
  F.~I.~Pelupessy, T.~Di Matteo and B.~Ciardi,
  arXiv:astro-ph/0703773.

\end{thebibliography}
\end{document}